\documentclass[aps,floatfix]{revtex4-2}
\usepackage{hyperref,hypcap,ae,tikz}
\usepackage{graphicx}
\usepackage{bm}
\usepackage{amsmath}
\usepackage[T1]{fontenc}
\usepackage{textcomp}
\usepackage{xcolor}
\usepackage{lmodern}
\usepackage[caption=false]{subfig}

\hypersetup{
	colorlinks=true,
	citecolor=blue,
    linkcolor=magenta,
	}
\begin{document}

\title{{\bf  Density perturbation and cosmological evolution in the presence of magnetic field in $f(R)$ gravity models}}

\author{\bf Samarjit Chakraborty$^{1,a}$}
\author{Sarbari Guha$^{1,b}$}

\affiliation{\bf $^1$ Department of Physics, St.Xavier's College (Autonomous), Kolkata 700016, India \\
$^{a}$ samarjitxxx@gmail.com \\ $^{b}$ srayguha@yahoo.com}

\maketitle
\section*{Abstract}
In this paper, we have investigated the density perturbations and cosmological evolution in the FLRW universe in presence of a cosmic magnetic field, which may be assumed to mimic primordial magnetic fields. Such magnetic fields have sufficient strength to influence galaxy formation and cluster dynamics, thereby leaving an imprint on the CMB anisotropies. We have considered the FLRW universe as a representative of the isotropic cosmological model in the 1+3 covariant formalism for $f(R)$ gravity. The propagation equations have been determined and analyzed, where we have assumed that the magnetic field is aligned uniformly along the $x$-direction, resulting in a diagonal shear tensor. Subsequently, the density perturbation evolution equations have been studied and the results have been interpreted. We have also indicated how these results change in the general relativistic case and briefly mentioned the expected change in higher-order gravity theories.

\bigskip

KEYWORDS: FLRW universe; magnetic field; modified gravity models; $f(R)$ gravity; perturbation.

\section{Introduction}

Study of density perturbation evolution in presence of primordial magnetic field (PMF) with higher curvature contributions in the gravitational actions is of considerable importance in determining the physical viability of a cosmological model in any theory of gravity. These PMFs in the early universe are expected to be the origin of the large-scale cosmic magnetic fields. The widespread existence of magnetic fields in the universe \cite{Kronberg:rpp1994}, possess sufficient strength to suggest that galaxy formation as well as cluster dynamics are, to some extent, influenced by magnetic forces. Although large-scale magnetic fields are not expected to survive an epoch of inflation, but such fields could also be generated even at the end of inflation or during the subsequent phase transitions \cite{TM:prd2000,TV}. Even though the origin of such magnetic fields may be debated upon \cite{TV1}, but their effects on the anisotropies present in the cosmic microwave background (CMB) due to both the electromagnetic and gravitational interactions, will provide better understanding of the early evolutionary phase of the universe.

Even though people initially thought that PMFs did not produce any relevant effect after recombination, later it was realized that PMFs might have significant role in several fundamental processes which occurred in the initial $ 10^5 $ years, e.g., it may have affected the big-bang
nucleosynthesis, the dynamics of some phase transitions, and baryogenesis \cite{Rub}. The hypercharge component of PMF has a net helicity (Chern-Simon number), which using the Abelian anomaly, may have been converted into baryons and leptons. Thus, PMF plays an important role in the production of the matter-antimatter asymmetry of the Universe \cite{GraRub}. Moreover, cosmic magnetic fields must exist everywhere in our universe. Recent astronomical studies have revealed that the magnetic fields are crucial for the origin and evolution of different astrophysical systems such as solar flares, fast radio burst, binary systems, Gamma-ray pulsars, magnetars, active galactic nuclei jets, galaxy-cluster evolution etc. \cite{Deng1,Deng2,Gao4,WangGao, Gao3}. Consequently different varieties of cosmological processes also can produce relatively high primordial magnetic fields of strength $ 10^{-10}-10^{-9} $ gauss at the redshift $ z\sim 5 $ \cite{GraRub}. Recently the effect of these PMFs on the CMB anisotropies on smaller scales than the mean-free-path of the CMB photons, have been investigated \cite{MIT:jcap2021}. The measurements of CMB anisotropies may be used to compare the predictions of a given cosmological model with the observed values to establish the viability of a chosen model of the universe.

The universe is in a state of accelerated expansion in the current epoch \cite{Reiss.et.al:AJ1998, Perlmutter.et.al:AJ1999}. This observed acceleration is usually addressed by either modifying the Einstein-Hilbert action, or by introducing a ``dark energy'' component in the energy momentum tensor of matter \cite{PR:rmp2003, Padmanabhan:PR2003, CST:ijmpd2006}. The simplest dark energy (DE) model which is consistent with the observations is the Lambda cold dark matter ($\Lambda$CDM) model. However, the extraordinary fine-tuning necessary for the value of the cosmological constant to be compatible with the observational results, led researchers to explore other models including the dynamical DE ones, as well as modified gravity models. The simple version of the modified gravity models is the $f(R)$ gravity model, where the Einstein-Hilbert action includes higher power terms of the scalar curvature $R$ \cite{Buchdahl:MNRAS1970, Starobinsky:PLB1980, Capozziello:IJMPD2002, CVMT:prd2004}. The first $f(R)$ gravity which passes solar tests was proposed in \cite{NO1:prd2003}. Several reviews of modified gravity models can be found in literature \cite{modgrav}.

In \cite{lee} the authors have linked the problem of dark energy to the issue of generation of PMF. They studied the PMF generation during the large scale structure formation of the universe by coupling the electromagnetic field to an evolving pseudo scalar field which accounts for the dark energy dynamics. Though the origin and composition of DE still remains unresolved, the observational data can be used to test the dynamic property of DE \cite{Hua}. The study of these dynamic properties can give us crucial information about the PMF and its interaction with the pseudo scalar field. Another interesting aspect of this scenario is that magnetic fields present in the early Universe are capable of generating and amplifying baryonic matter density perturbations through the Lorentz force and the presence of gravitational coupling enables these density perturbations to propagate to the dark matter fluid as well, leading to a significant change not only in the distribution of density fluctuations but also in the structure formation \cite{fedeli}.

In $1967$, Field and Shepley \cite{FS:Ap&SS1968} studied the density perturbations in cosmological models and demonstrated that density perturbations with long wavelengths are unstable for all time. In $ 1984 $, Kodama and Sasaki reviewed and reformulated extensively the linear perturbation theory of spatially homogeneous and isotropic universes \cite{KS:ptps1984}. Souradeep and Sahni \cite{SS:mpla1992} studied the contributions to the observed large scale anisotropy in the cosmic microwave background radiation (CMBR) arising from gravitational waves and adiabatic density perturbations generated by a common inflationary mechanism in the early universe. They also found that in the inflationary models predicting a power law primordial spectra, i.e., $|\delta_k|^2 \propto k^n$, the relative contributions to the quadrupole anisotropy from gravity waves and scalar density perturbations, depends on $ n $.

Tsagas and Maartens \cite{TM:prd2000} performed a fully relativistic treatment on all the effects on the growth of density perturbations, rotational instabilities and anisotropic deformation in the density distribution, incorporating the magneto-curvature coupling that arises in a relativistic approach. Marklund et al. \cite{MDB:prd(R)2000} investigated the dynamics of electromagnetic fields in an almost-Friedmann-Robertson-Walker universe using the covariant and gauge-invariant approach of Ellis and Bruni \cite{EB:prd1989}. They have shown that the coupling between gravitational waves and a weak magnetic test field can generate electromagnetic waves. Barrow et al. \cite{BMT:phys-rept2007} reviewed the spacetime dynamics in presence of large-scale electromagnetic fields, and then considered the effects of the magnetic component on perturbations to a spatially homogeneous and isotropic universe. Z. F. Gao et al also studied the other properties of gravity, such as the scattering of Dirac spinors and the scalar fields, which can also be very useful in the study of PMFs \cite{Gao1,Gao2}.

Linear theory of perturbations in a FLRW universe with a cosmological constant was studied by Vale and Lemos \cite{VL:mnras2001}, and the equations for the evolution of the perturbations were derived in the fully relativistic case. In \cite{LBlanc} the Einstein--Maxwell field equations for orthogonal Bianchi $VI_{0}$ cosmologies with a $ \gamma $-law perfect fluid and a pure, homogeneous source-free magnetic field, were expressed as an autonomous differential equation in terms of expansion-normalized variables. Pereira et al. \cite{PPU:jcap2007} described the theory of cosmological perturbations around a homogeneous and anisotropic universe of the Bianchi I type. In \cite{GCP:jcap2007} the authors extended the standard theory of cosmological perturbations to homogeneous but anisotropic universes using the case of a Bianchi I model, with a residual isotropy between two spatial dimensions, which is undergoing complete isotropization at the onset of inflation. Recently in \cite{GM:epjc2019} the authors examined the effect of a magnetic field on the growth of cosmological perturbations by developing a mathematical treatment in which a perfect fluid and a uniform magnetic field evolve together in a Bianchi I universe.

In the papers \cite{ML:cqg1992,miedema:prd1994}, the evolution equations for small perturbations in the metric, energy density and material velocity for an anisotropic viscous Bianchi I universe were studied. In \cite{DFP:jhep2008} the authors examined the growth of perturbations in an expanding Bianchi type-I metric dominated by dust and a cosmological constant. Barrow and Hervik \cite{Barrow} investigated the Bianchi type I brane-worlds with a pure magnetic field filled with a perfect fluid to find new properties of such brane-worlds. Clarkson et al. \cite{CCQ:prd2001} examined the qualitative properties of the class of spatially homogeneous Bianchi type-$VI_{o}$ cosmological models containing a perfect fluid with a linear equation of state, a scalar field with an exponential potential and a uniform cosmic magnetic field, using dynamical systems techniques.

Watanabe et al. \cite{WKS:ptp2010} investigated the statistical nature of primordial fluctuations from an anisotropic inflation conceptualized by a vector field coupled to an inflaton. They found the cross correlation between the curvature perturbations and gravitational waves, which never occurs in conventional inflation. Bamba et al. \cite{bamba} demonstrated that the subhorizon approximation in cosmological perturbation theory is not always a good approximation at least in the case of the $k$-essence model of dark energy. They showed that the sound speed in the these models has a huge influence on the time evolution of the matter density perturbation, and future observations may clarify the differences between the $ \Lambda $CDM model and the $k$-essence model.

Ellis and MacCallum \cite{EM:cmp1969} provided a detailed study of homogeneous cosmological models. In \cite{WH:cqg1989} Wainwright and Hsu considered the problem of describing the asymptotic states of orthogonal spatially homogeneous cosmologies, near the big bang and late times. The paper \cite{MEEM:cqg2000} describes the partially locally rotationally symmetric (LRS) perfect fluid cosmologies.
In 1997, Elst and Uggla \cite{EU:cqg1997} explicitly discussed the dynamical equations of an extended $ 1+3 $ orthonormal frame approach in the relativistic description of spacetime geometries. In the paper \cite{BDFK:cqg2012} the authors studied the growth of density perturbations in Kantowski-Sachs cosmologies with a positive cosmological constant, using the $ 1+3 $ and $ 1+1+2 $ covariant formalisms.

In their paper \cite{LCS:pdu2017}, the authors investigated the cosmological dynamics of a uniform magnetic field in the Orthogonal Spatially
Homogeneous (OSH) Anisotropic Bianchi I universe with viable $f (R)$ gravity models. They found that there exist an additional fixed point before the emergence of the standard matter epoch, which shows the existence of the primordial magnetic fields and the anisotropy of spacetime before the universe expands to assume isotropic geometry. In \cite{NO2:prd2006,CNOT:plb2006,CENOSZ:prd2008} the authors formulated several versions of $f(R)$ gravity and its modified form consistent with realistic cosmology, which were compatible with solar system tests. Matsumoto \cite{Matsumoto:prd2013} investigated the oscillating solutions of the matter density perturbation in $f(R)$ gravity models using appropriate approximations. It was concluded that the condition $f_{RR}/f_{RRR}>0$ helps us to generate a model which yields almost equivalent background evolution to the $\Lambda$CDM model, but yields different evolution of the matter density perturbation compared to the $\Lambda$CDM model.

Amendola and Tsujikawa \cite{AT:plb2008} identified the class of $ f(R) $ dark energy models which have a matter dominated epoch followed by a late-time acceleration. The representative models that satisfy both cosmological and local gravity constraints take the asymptotic form $m(r) = C(-r-1)^p$ with $p >1$ as $r$ approaches $-1$ where the deviation from a $ \Lambda $CDM model $(f = R -\Lambda)$ is quantified by the function $m = Rf_{,RR}/f_{,R}$. In \cite{leach} the authors obtained the equations governing the shear evolution in Bianchi spacetimes for general $ f(R) $ theories of gravity.
The paper \cite{MRSTepjc2018} investigated the past evolution of an anisotropic Bianchi I universe in $ R+R^2 $ gravity.

An analytic solution for density perturbations in the matter component during the matter dominated stage was obtained in the paper \cite{MSY:ijmpd2009} in terms of hypergeometric functions for a class of viable $f(R)$ models in which the deviation from Einstein gravity decreases as an inverse power of the Ricci scalar. An analytical expression for the matter transfer function was also obtained at scales much smaller than the present Hubble scale. The authors in \cite{SHS:prd2007} studied the evolution of linear cosmological perturbations in $ f(R) $ models of accelerated expansion with minimally coupled matter in fourth order gravitational dynamics. These models predict testable set of linear phenomena which are accessible with current and future data giving stringent tests of general relativity on cosmological scales.

In the paper \cite{MHT:epjc2012}, the authors obtained field equations in a special $ f(R) $ gravity in the presence of external, linear/nonlinear electromagnetic fields. The paper \cite{CDT:prd2008} gave a detailed mathematically well defined presentation of the covariant and gauge invariant theory of scalar perturbations of a FLRW universe for fourth order gravity, where the matter is described by a perfect fluid with a barotropic equation of state. They obtained exact solutions of the perturbations equations for scales much bigger than the Hubble radius and discussed their various interesting features.
In \cite{TUT:prd2008} the authors studied of matter density perturbations in both metric and Palatini formalisms. Considering a general function of the Ricci scalar $ R $, they derived the equation of matter density perturbations in each case, in a number of gauges, demonstrating that for viable $ f(R) $  models that satisfy cosmological and local gravity constraints (LGC), matter perturbation equations derived under a subhorizon approximation are valid even for super-Hubble scales provided the oscillating mode (scalaron) does not dominate over the matter-induced mode. Scalar perturbations of the metric for nonlinear $ f(R) $ models have been studied by some authors in the case of the universe at the late stage of its evolution \cite{ENZ:epjc2014}. A review of the investigations on the structure formation in $ f(R) $-gravity based on the Covariant and Gauge Invariant approach to scalar perturbations is available in  \cite{Carloni:TOAJ2010}.

Maartens and Taylor \cite{maartens} examined the kinematic and dynamic properties of fluid spacetimes in higher order gravity, and extended the general equations of Ehlers and Ellis governing relativistic fluid dynamics from general relativity (GR) to the higher order theory.
In \cite{rippl}, the authors extended the equations of relativistic fluid dynamics given by Ehlers and Ellis for GR, and by Maartens and Taylor for quadratic theories, to generalised $f(R)$ theories of gravity. The dynamics of OSH Bianchi cosmologies was studied by Goheer et al. in $ R^n $-gravity \cite{goheer}.

In \cite{SNAC-DM:epjc2020} the cosmological scalar perturbations of standard matter are investigated in the context of extended teleparallel $ f(T) $ gravity theories using the $ 1+3 $ covariant formalism. The authors concluded that $ f(T) $ models contain a richer set of observationally viable structure growth scenarios that can be tested against observational data and are consistent with currently known features of the large-scale structure power spectrum in the GR and $\Lambda$CDM limits. De Felice and Suyama \cite{DeFeliceSuyama:ptp2011} derived the equation of matter density perturbations on sub-horizon scales around a flat FLRW background for the general Lagrangian density $f(R, G)$ that is a function of a Ricci scalar $ R $ and a Gauss-Bonnet term $ G $ demonstrating the effective gravitational constant generically scales as distance squared at small distances.

Harko et al. \cite{harko:prd2011} proposed and discussed in details the $ f(R,T) $ modified theory of gravity, where the gravitational Lagrangian is given by an arbitrary function of the Ricci scalar $ R $ and of the trace of the stress-energy tensor $ R $.
Subsequently in the paper \cite{AC-DHRS-Z:prd2013}, the authors studied the evolution of scalar cosmological perturbations
in the metric formalism in the context of $ f(R,T) $ theories of gravity. They found the complete set of differential equations for the matter density perturbations, where in the case of sub-Hubble modes, the density contrast evolution reduces to a second-order equation. In \cite{PJ:ijgmmp2018}, a class of spatially homogeneous and anisotropic Bianchi-V massive string models have been studied in a modified $ f(R,T) $-theory of gravity in the presence of a magnetic field. They found that the expanding solution is stable against the perturbation with respect to anisotropic spatial direction.

This work is an extension of the previous works in $f(R)$ theory to examine the effect of magnetic field in the evolution of density perturbation.
We find that the effect of magnetic field is expected as it behaves like an effective magnetized fluid in the matter sector of the action. The paper is organized as follows: Section II describes the background equations in 1+3 covariant formalisms in presence of a cosmic magnetic field. The linearized equations are given in Section III, followed by the subhorizon approximation and ultrarelativistic approximation in Section IV. In Section V, we have presented the evolution equations in the general relativistic case and other higher
order theories. We end with some discussions of the results obtained and our conclusive statements in Section VI.

\section{Background equations in 1+3 covariant formalisms in presence of magnetic field in $f(R)$ theory}

Let us consider the action for $f(R)$ gravity model in the form \cite{Matsumoto:prd2013}
\begin{align}
S=\frac{1}{2 \kappa ^2} \int d^4 x \sqrt{-g} \left [ R+f(R) \right ]+S_\mathrm{matter},
\label{10}
\end{align}
where $f(R)$ is an arbitrary function of the scalar curvature $R$, which represents the deviation from Einstein gravity. Here $\kappa^2 = 8\pi G = 1/M^2_{pl}$, $G$ being the gravitational constant, $M_{pl}$ is the reduced Planck mass, and $S_\mathrm{matter}$ is the matter part of the action. We want to point out that contrary to the popularly used functional form (i.e., $f(R)$), we have used this decomposed version $R+f(R)$ in the action. This choice of the $f(R)$ function helps us to keep track of the curvature terms in the equations, and we can quickly derive the corresponding equations for GR by simply setting $ f(R)=0 $. This is also in conformity with the functional form of the action in the Starobinsky model where the function is $ R+\alpha R^2 $. According to our choice, $ f(R)= \alpha R^2 $. This is only for mathematical convenience, the physical interpretation does not get affected by it.  

Varying the above action with respect to the metric $g_{ab}$, and after some manipulations (see standard literature on $f(R)$ gravity), the gravitational field equation can be reduced to the form:
\begin{equation}
\label{eq:einstScTneff}
R_{ab}=\frac{1}{(1+f_{R})}\left(\frac{1}{2}g_{ab}(R+f(R))-(g_{ab}g^{cd}-g_a^cg_b^d)S_{cd}+T_{ab}\right)
\end{equation}
where $S_{ab}=\nabla_a\nabla_b (1+f_{R})$\,.

Let us define the energy-momentum tensor $T_{ab}$ as:
\begin{equation}\label{eq1}
T_{ab}=\rho u_au_b+ph_{ab}+Q_au_b+Q_bu_a+\Pi_{ab}\, ,
\end{equation}
where $h_{ab}=g_{ab}+u_au_b$ is the induced metric, which is associated with the spatial hypersurface. The four-velocity given by $u^a=(\,1\,,~0\,,~0\,,~0\,)$ is orthogonal to $h_{ab}$ (i.e., $h_{ab}u^a=0$)\,. Moreover $Q_a$ denotes the energy flux such that $Q_a u^a=0$, and the symmetric trace-free anisotropic pressure is denoted by $\Pi_{ab}$. Here $\Pi_{a}^a=0 , \, \Pi_{ab}u^a=0$ are all relative to $u^a$ \cite{leach}. Now let us consider an uniform magnetic field, so that electromagnetic contributions will be present in the energy momentum tensor. Consequently the energy momentum tensor in \eqref{eq1} can be decomposed into two parts as the following:
\begin{equation}
T_{ab}=T_{ab}^{PF}+T_{ab}^{EM},
\end{equation}
where $T_{ab}^{PF}$ is the energy-momentum tensor of the uncharged matter assumed to be represented by a perfect fluid, and therefore it can be expressed using $ u^{a} $ as the following:
\begin{equation}
T_{ab}^{PF}=\rho_m u_au_b+p_mh_{ab}\,.
\end{equation}
In order to know the full functional ecpression of the energy momentum tensor we must also have the expression of the contribution responsible for the electromagnetic interactions (in this case an uniform magnetic field). Consequently  \indent $T_{ab}^{EM}$ is the energy-momentum tensor for the Maxwell field, which is given by
\begin{eqnarray}
T_{ab}^{EM}&=&F_{ac}F^{c}_{~b}-\frac{1}{4}g_{ab}F_{cd}F^{cd},
\end{eqnarray}
where $F_{ab}$ is the the field strength and defined by the expression \eqref{eq2}:
\begin{equation}\label{eq2}
F_{ab}=\frac{1}{2}\,u_{[a}E_{b]}+\eta_{abcd}H^cu^d.
\end{equation}
Here $E_a$ and $H_a$ are the electric and magnetic fields respectively. In this work we will assume the energy-momentum tensor of the Maxwell field to be a pure magnetic case. For simplicity of calculations, let us assume that the uniform magnetic field is aligned in the $x$-direction only. Therefore we can take the magnetic fields with components as $H_a=(\,0\,,~\tilde H\,,~0\,,~0\,)$\, \cite{LBlanc}\,.

We can treat the energy-momentum tensor of the Maxwell fields as analogous to the case of a perfect fluid \cite{Barrow}, having an expression described by \eqref{eq3}:
\begin{equation}\label{eq3}
T_{ab}^{EM}=\rho_{EM}u_au_b+p_{EM}h_{ab}+\Pi_{ab},
\end{equation}
where the energy density of the electromagnetic field is $\rho_{EM}=\frac{1}{2}\tilde H^2$\,, the pressure is given by $p_{EM}=\frac{1}{6}\tilde H^2$\, and $\Pi_{ab}$ is given by
\begin{equation}
\Pi_{ab}=\frac{1}{3}H^2h_{ab}-H_aH_b \,.
\end{equation}
The quantity $H^2\equiv H_aH^a = \tilde H^2 $ is the magnitude of magnetic fields expressed in terms of the field strength.

\indent As the energy-momentum tensor of the Maxwell field is trace-free ($g^{ab}T_{ab}^{EM}=T^{EM}=0$)\,, the energy density $\rho$ and the pressure $p$ can be decomposed into the matter and the magnetic parts as the following:
\begin{equation}
\rho = \rho_{PF} + \rho_{EM}\,,\quad \rho_{PF}=\rho_m\,,\quad \rho_{EM}=\frac{1}{2}\,\tilde H^2\,,\qquad\quad p = p_{PF} + p_{EM}\,,\quad p_{PF}=p_m\,,\quad p_{EM}=\frac{1}{6}\tilde H^2\,.
\end{equation}

Now we will describe the propagation equations \cite{WH:cqg1989, EM:cmp1969, EU:cqg1997, MEEM:cqg2000} for $f(R)$ gravity in  $1+3$ covariant formalism. The  $1+3$ orthogonally spatially homogeneous formalism uses the fluid velocity as time-like vector fields which orthogonalize along with the spatial vector fields. We will be assuming a LRS geometry on the spacetime, where the spatial vector fields span the space-like hypersurface along with a preferred spatial direction. As there is local rotational symmetry, the spatial vectors remain invariant under the rotation of the preferred spatial axis (the $x$-direction, parallel to our chosen magnetic field). Using our previous expression for the Ricci tensor in Eq.(\ref{eq:einstScTneff})\,, we can always split \cite{leach,rippl} $R_{ab}$ into the following forms:

\begin{eqnarray}
R&=&(1+f_{R})^{-1}(T+2f-3S),\\
R_{ab}u^au^b&=&(1+f_{R})^{-1}(T_{ab}u^au^b-\frac{1}{2}f+h^{ab}S_{ab}),\\
R_{ab}u^ah_c^b&=&(1+f_{R})^{-1}(S_{ab}u^ah_c^b-q_c),\\
R_{ab}h_c^ah_d^b&=&(1+f_{R})^{-1}\left(\Pi_{cd}-(p+\frac{1}{2}f+S)h_{cd}+S_{ab}h_c^ah_d^b\right)\,,
\end{eqnarray}
and similarly for the $S_{ab}$ the same decomposition can be done as
\begin{eqnarray}
S&=&-f_{R}'(\ddot{R}+\Theta\dot{R})-f_{R}''\dot{R}^2,\\
S_{ab}u^au^b&=&f_{R}'\ddot{R}+f_{R}''\dot{R}^2,\\
S_{ab}h^{ab}&=&-f_{R}'\Theta\dot{R}\,. \label{17}
\end{eqnarray}
Explicit study of the $1+3$ covariant analysis in the $f(R)$ gravity models can be found in Refs.\cite{leach} and \cite{goheer}. Following the above analysis (using \eqref{17}), we obtain the Raychaudhuri equation as given in \eqref{raychua}:
\begin{eqnarray}\label{raychua}
&& \dot{\Theta}+\frac{1}{3}\Theta^2+2\Sigma^2+\frac{1}{(1+f_{R})}\left(\rho-\frac{1}{2}f+h^{ab}S_{ab}\right) = 0 \, \Rightarrow
\dot{\Theta}+\frac{1}{3}\Theta^2+2\Sigma^2+\frac{1}{(1+f_{R})}\left(\rho-\frac{1}{2}f - f_{R}'\Theta\dot{R} \right)=0\,,
\end{eqnarray}
and  subsequently we take the first integral of the above equation \eqref{raychua} to get the Friedmann equation in \eqref{friedmann}:
\begin{eqnarray}\label{friedmann}
&&\frac13\Theta^2 - \Sigma^2 - \frac{1}{(1+f_{R})}\left(\rho + 3p + f - 3S + 2h^{ab}S_{ab} \right) = 0
\nonumber\\
&& \Rightarrow \frac13\Theta^2 - \Sigma^2 - \frac{1}{(1+f_{R})}\left(\rho + \frac{1}{2}((1+f_{R})R - f) - f_{R}'\Theta\dot{R} \right) = 0 .
\end{eqnarray}
A very important thing to note here is that both in \eqref{raychua} and \eqref{friedmann}, the effect of the magnetic field is present via the parameters $ \rho $ and $ p $ of the effective cosmic fluid, influencing the evolution of the universe via the rate of expansion and the rate of Ricci scalar. Moreover the shear propagation equation is given as follows:
\begin{equation}\label{sheardot}
\dot{\Sigma}_{ab}+\Theta\Sigma_{ab}=\frac{1}{(1+f_{R})}\left(
\pi_{ab}-f_{R}'\dot{R}\,\Sigma_{ab}\right),
\end{equation}
where $\Theta\equiv \Theta_{ab}h^{ab}$ is the rate of volume expansion, $\Sigma^2\equiv \frac{1}{2}\Sigma_{ab}\Sigma^{ab}$ is magnitude of the shear tensor
$\Sigma_{ab}$ (with $\Sigma_{ab}=\Theta_{ab}-\frac{1}{3}h_{ab}\Theta$\,, $\Sigma_a^a=0$ and $\Sigma_{ab}u^a=0$)\,.

We know that during structure formation, the shear plays a very important role. Therefore the evolution of the shear tensor is very crucial, and the effect of magnetic field in the evolution of the shear tensor is very much evident from the above equation \eqref{sheardot}. The magnetic field is affecting the structure formation via the ansitropic pressure like term $\Pi_{ab} $, which directly depends on the magnetic field strength.

Now we decompose the tetrad field into the orthonormal frame according to \cite{EU:cqg1997, MEEM:cqg2000}\,. Here we align the frame in such a way so that
$ \mathbf{e}_{0}=u=\frac{\partial}{\partial t} $, where $ u $ is the fluid vector field, and the spatial triad $ \lbrace\mathbf{e}_{\alpha}\rbrace $ spans the tangent space. The commutation relation among the vectors is
\begin{equation}
[\mathbf{e}_{a},\mathbf{e}_{b}]=\gamma^{c}_{ab}\mathbf{e}_{c}.
\end{equation}
Here the purely spatial components, $\gamma^{c}_{ab} $, are being decomposed into an object $ a_{a} $, and a symmetric object $ n_{ab} $ as follows:
\begin{equation}
\gamma^{a}_{bc}=2a_{[b}\delta^{a}_{c]}+\epsilon_{bcd}n^{da},
\end{equation}
where $ \epsilon_{abc} $ is the totally antisymmetric $ 3 $-D permutation tensor. It is to be noted that we have the algebraic constraint $ n^{a}_{b}\Sigma^{b}_{c}=\Sigma^{a}_{b}n^{b}_{c} $, and using this freedom \cite{LBlanc, CCQ:prd2001} we can simultaneously diagonalize $ \Sigma_{ab} $ and $ n_{ab} $ as the following:
\begin{equation}
\Sigma_{ab}=\text{diag}(\,\Sigma_{11}\,,~\Sigma_{22}\,,~\Sigma_{33}\,)
\equiv \text{diag}(\,\Sigma_{1}\,,~\Sigma_{2}\,,~\Sigma_{3}\,),\,\,\,\,\,\, \&  \,\,\,\,\,\,\,  n_{ab}= \text{diag}(\,n_{1}, n_{2}, n_{3}\,).
\end{equation}
There is also another constraint, which is: $ n_{ab}H^{b}=0 $. For the sake of simplicity, we can always align the direction of the magnetic field along that of the shear eigenvector.
Therefore our shear propagation equation gets diagonalized and reduces to a vector equation from a tensor equation. This can be seen in equation \eqref{sheardia}.
\begin{eqnarray}\label{sheardia}
\dot{\Sigma}_{a}+\Theta\Sigma_{a}=\frac{1}{(1+f_{R})}
\left(\Pi_{a}-f_{R}'\dot{R}\,\Sigma_{a}\right),
\end{eqnarray}
where the analogous anisotropic magnetic pressure becomes $\Pi_a \equiv \Pi_{aa}$, and $\Pi_{aa}$ is the diagonal elements of $\Pi_{ab}$ tensor\,.

Using the conservation of energy-momentum tensor along with the source-free of Maxwell field we obtain \cite{LBlanc} the propagation equation of matter parts described below (\eqref{rhodot} and \eqref{hdot}):
\begin{eqnarray}\label{rhodot}
\dot{\rho_m}&=&-(1+w)\rho_m\Theta,
\end{eqnarray}
\begin{eqnarray}\label{hdot}
\dot{\tilde H}&=&-\frac{2}{3}\Theta \tilde H + \Sigma_{1}\tilde H = -\frac{2}{3}\Theta \tilde H - 2(\Sigma_2 + \Sigma_3)\tilde H\,.
\end{eqnarray}
In order to emphasize the effect of magnetic field on the entire dynamics of the universe we present another useful relation among the Ricci curvature and the Raychaudhuri variables. With the help of the Raychaudhuri equation \eqref{raychua} and the Freidmann equations \eqref{friedmann}, we can obtain the following relation in \eqref{R-thetadot}:
\begin{eqnarray}\label{R-thetadot}
R &=&2\dot{\Theta} + \frac{4}{3}\Theta^2 + 2\Sigma^2\,.
\end{eqnarray}%
Note that magnetic field is not visible in the equation, but from the Raychaudhuri equation it is evident that the term $ \dot{\Theta} $ is being influenced by the magnetic field, i.e., not only the structure formation but even the rate of expansion of the universe is also influenced by the magnetic field. Therefore the role of magnetic field in the evolution of the universe cannot be understated. It is to be noted that we are using the isotropic FLRW spacetime as the background with an anisotropic magnetic field in order to closely mimic an almost isotropic magnetized universe without magnetocurvature coupling. We are not addressing specifically the exact anisotropic magnetized solutions of Einstein-Maxwell system. This is done for mathematical convenience where we have attributed the anisotropy not to the background spacetime but to the magnetic field. Therefore, this present study deals with isotropic spacetimes having magnetic field anisotropy in $ f(R) $ theories, which may lead to anisotropic effects. We do so because we want to study the effects of magnetized background on the density perturbations in such cases. In the subsequent sections we have shown how the presence of such magnetic field affects the cosmological evolution, and how it affects the density perturbation.

\section{Linearized equations in 1+3 covariant formalisms in presence of magnetic field}
In order to see the effect of magnetic fields in the density perturbation we will consider the spatially flat Friedmann-Lemaitre-Robertson-Walker (FLRW) metric which is given in the form \eqref{frd}:
\begin{equation}\label{frd}
ds^2 = a^2(\eta) (d\eta^2-\sum_{i=1}^3 dx^i dx^i).
\end{equation}
Consequently we express the Friedmann-Lemaitre equations in the following way:
\begin{align}
\frac{3\mathcal{H}'}{a^2}(1+f_{R})-\frac{1}{2}(R+f)
-\frac{3\mathcal{H}}{a^2}f_{R}' &=- \kappa ^2 \rho
\label{FL00}, \\
\frac{1}{a^2}(\mathcal{H}'+2\mathcal{H}^2)(1+f_{R})
-\frac{1}{2}(R+f)-\frac{1}{a^2}(\mathcal{H} f_{R}'+f_{R}'') &= \kappa^2 w \rho
\label{FLii},
\end{align}
where the Ricci tensor is given by $R = 6a^{-2} (\mathcal{H}'+\mathcal{H}^2)$, and the derivative of the $ f(R) $ function with respect to $ R $ is being expressed as $f_{R} \equiv df(R)/dR$. Here the prime represents the differentiation with respect to conformal time $\eta$. Also the energy density of the matter is given by $\rho$, which is coming from the variation of $S_\mathrm{matter}$, and $w$ is the equation of state parameter defined by $w=p/ \rho$. The Hubble rate with respect to conformal time is expressed by $\mathcal{H}$ which is defined by the relation $\mathcal{H}\equiv a'/a$.
The equation of continuity can be written as follows:
\begin{equation}
\rho'+3(1+w)\mathcal{H}\rho = 0.
\label{15}
\end{equation}
Finally the Friedmann-Lemaitre equations along with the continuity equation, form the background equations of the Universe.

In order to find the perturbation equations for matter density we begin with the FLRW metric in the Newtonian gauge\eqref{nfrw}:
\begin{equation}\label{nfrw}
ds^2 = a^2(\eta) [(1+2\Phi) d\eta^2- (1+2\Psi)\sum_{i=1}^3 dx^i dx^i ].
\end{equation}
Moreover, the elements of the linearized Einstein equation in the Fourier space (i.e. $(00)$, $(ii)$, $(0i)=(i0)$, and $(ij)$, $i\neq j$ components) are represented respectively as follows:
\begin{align}
(1+f_{R}) \{ &-k^2(\Phi+\Psi)-3\mathcal{H}(\Phi'+\Psi')+(3\mathcal{H}'
-6\mathcal{H}^2)\Phi-3\mathcal{H}'\Psi \} \nonumber \\
&+f'_{R}(-9\mathcal{H}\Phi+3\mathcal{H}\Psi-3\Psi')\,=\,\kappa ^2 \rho a^2 \delta .
\end{align}\label{R00}
This equation involves only the first derivative in conformal time. Next we have
\begin{align}
(1+f_{R})\{ \Phi''+&\Psi''+3\mathcal{H}(\Phi'+\Psi')
+3\mathcal{H}'\Phi+(\mathcal{H}'+ 2\mathcal{H}^2)\Psi \} \nonumber \\
+&f'_{R}(3\mathcal{H}\Phi-\mathcal{H}\Psi+3\Phi') + f''_{R}(3\Phi-\Psi)\,=\,
c_{s}^{2} \kappa ^2 \rho a^2 \delta . \label{rr}
\end{align}
Interestingly this equation involves second derivatives in conformal time along with the first derivatives. Then we have
\begin{align}
(1+f_{R})\{ \Phi'&+\Psi'+\mathcal{H}(\Phi+\Psi) \}+f'_{R}(2\Phi-\Psi)\,=\,
- \kappa ^2 \rho a^2 (1+w) v .
\end{align}\label{R0i}
This equation also involves first derivatives in conformal time, and finally we have the expression involving a double derivative in $ R $ (unlike the previous three where we have $ f_{R} $):
\begin{align}
\Phi-\Psi - \frac{2f_{RR}}{a^2(1+f_{R})}& \{ 3\Psi''+6(\mathcal{H}'+\mathcal{H}^2)\Phi
+3\mathcal{H}(\Phi'+3\Psi')-k^2(\Phi-2\Psi) \} = 0
\label{Rij}.
\end{align}
Here, the sound speed is defined by $c_\mathrm{s}^2 \equiv \delta p / \delta \rho$, and the matter density perturbation is expressed as
$\delta \equiv \delta \rho / \rho.$
Next we perturb the equation of continuity, $\nabla _\mu T^{\mu \nu}=0$, and  we have the following relations:
\begin{equation}
3\Psi'(1+w)-\delta'+3 \mathcal{H}(w-c_\mathrm{s}^2)\delta +k^2(1+w)v=0 ,
\label{T0}
\end{equation}
and
\begin{equation}
\Phi+\frac{c_{s}^2}{1+w}\delta+v'+\mathcal{H}v(1-3w)=0.
\label{Ti}
\end{equation}

\section{Subhorizon approximation and ultrarelativistic approximation in presence of magnetic field}

In this section we derive the differential equation of the matter density perturbation in the sub-horizon approximation. Firstly, let us assume the quasi-static approximation, i.e., $\partial/\partial \eta \sim \mathcal{H}$. We also need to take into consideration the small scale approximation, which is given by $\mathcal{H} \ll k$.
Therefore in the subhorizon approximation we combine these two conditions to have $\partial/\partial \eta \sim \mathcal{H} \ll k$. Next we apply sub-horizon approximation to the linearized Einstein equations in Fourier space to deduce the equation of the matter density perturbation. Necessary precautions regarding this problem was discussed in details in the paper \cite{Matsumoto:prd2013}.


Unlike in the case of the exact calculations, the procedure to derive the equation of the matter density
perturbation is not unique in the approximated calculations. We follow the process recommended in \cite{Matsumoto:prd2013}, and using the limit $\vert f_\mathrm{RR} \vert k^2/a^2 \ll 1$, we get the evolution equation of the density perturbation as the following:

\begin{align}
\dfrac{\delta \rho''}{\rho}+\left(\mathcal{H}-\dfrac{2\rho'}{\rho}\right)\left[\left(\dfrac{\delta \rho_{B}}{\rho}\right)'+\left(\dfrac{\delta \rho_{m}}{\rho}\right)'\right]-\dfrac{\delta \rho}{\rho}\left(\dfrac{\rho''}{\rho}+\dfrac{3a^2 \Omega_{m}}{2(1+f_{R})}\right)=0 .
\end{align}
The above equation can be decomposed into the form:
\begin{align}\label{dense}
\left(\dfrac{\delta \rho_{m}}{\rho}\right)''+ \mathcal{H} \left(\dfrac{\delta \rho_{m}}{\rho}\right)'-\dfrac{3a^2 \Omega_{m}}{2(1+f_{R})}\dfrac{\delta \rho_{m}}{\rho}+\mathbf{O_{B}}=0 .
\end{align}

Now if we take the approximation of a weak magnetic field compared to the matter contribution, i.e. $ \rho_{m}\gg\rho_{B} $, then the total energy density can be approximated as $ \rho\sim \rho_{m} $. This transforms the above evolution equation in to the following expression:
\begin{equation}\label{delh}
\delta_{m} '' + \mathcal{H} \delta_{m} ' - \frac{3a^2 \Omega_{\mathrm m}}{2(1+f_R)} \delta_{m} + \mathbf{O_{B}} = 0,
\end{equation}
where $\mathbf{O_{B}}  $ is given by
\begin{align}
\mathbf{O_{B}}\equiv \dfrac{\tilde{H}}{\rho}\delta\tilde{H}''+ \dfrac{\delta\tilde{H}'}{\rho}\left[2\tilde{H}'+\left(\mathcal{H}-\dfrac{2\rho'}{\rho}\right)\tilde{H}\right]+\dfrac{\delta \tilde{H}}{\rho}\Bigg[\tilde{H}''+\left(\tilde{H}'-\dfrac{\tilde{H}\rho'}{\rho}\right)\left(\mathcal{H}-\dfrac{2\rho'}{\rho}\right)-\tilde{H}\left(\dfrac{\rho''}{\rho}+\dfrac{3a^2 \Omega_{m}}{2(1+f_{R})}\right)\Bigg].
\end{align}

The expression in \eqref{delh} represents the density perturbation equation in a weak magnetic field. The higher contribution from the magnetic field is enclosed in the term $\mathbf{O_{B}}  $. We note that in absence of the magnetic field our result in equation \eqref{delh} reduces to the density perturbation equation obtained in \cite{Matsumoto:prd2013}, thereby extending their results to a universe dominated by a magnetic field.

Similarly, when $\vert f_\mathrm{RR} \vert k^2/a^2 \gg 1$, we have the following equation by using the sub-horizon
approximation:
\begin{align}
\delta_{m} '' + \mathcal{H} \delta_{m} ' - \frac{2a^2 \Omega_{\mathrm m}}{(1+f_R)} \delta_{m}+ \tilde{\mathbf{O}}_{B}= 0,
\label{302}
\end{align}
where similarly the magnetic contribution $\tilde{\mathbf{O}}_{B} $ is given by \eqref{to:2}

\begin{align}
\tilde{\mathbf{O}}_{B} \equiv \dfrac{\tilde{H}}{\rho}\delta\tilde{H}''+ \dfrac{\delta\tilde{H}'}{\rho}\left[2\tilde{H}'+\left(\mathcal{H}-\dfrac{2\rho'}{\rho}\right)\tilde{H}\right]+\dfrac{\delta \tilde{H}}{\rho}\Bigg[\tilde{H}''+\left(\tilde{H}'-\dfrac{\tilde{H}\rho'}{\rho}\right)\left(\mathcal{H}-\dfrac{2\rho'}{\rho}\right)-\tilde{H}\left(\dfrac{\rho''}{\rho}+\dfrac{2a^2 \Omega_{m}}{(1+f_{R})}\right)\Bigg] . \label{to:2}
\end{align}

Therefore in both the regimes we recover the expression of density perturbation equation for a non magnetic case discussed in \cite{Matsumoto:prd2013}. It may be noted that both $\mathbf{O_{B}}  $, $ \tilde{\mathbf{O}}_{B} $ are of second order in $ \delta \tilde{H} $ and depends inversely on the total energy density, i.e., larger is the matter contribution, smaller the magnetic contribution. It is as if the magnetic contributions are evolving side by side depending on the fluctuations in the magnetic field strength, although one must notice that these are not completely independent of the matter part as there are coupling terms present in both the cases.

We know that oscillating solutions cannot be obtained by sub-horizon approximation, and therefore we employ the ultra-relativistic approximation, i.e. $\partial/\partial \eta \sim k$, and small scale approximation $\mathcal{H} \ll k$, instead. We will only briefly describe how the ultra-relativistic  approximation can be done in the presence of a uniform magnetic field. In short, we say that to obtain the equation of $\delta$, we first assume a relation $\Phi \approx -\Psi$ from the leading terms of Eq. \eqref{rr}, and using Eq. \eqref{Rij} we arrive at the following:
\begin{align}
\Psi + \frac{3 f_{RR}}{a^2(1+f_R)}(\Psi'' + k^2 \Psi) \approx 0.
\label{303}
\end{align}
Now if we assume the condition $\vert f_{RR} \vert k^2/a^2 \ll 1$, then the $f(R)$ gravity models which satisfy such a condition do not possess oscillating solutions of $\Psi$ and $\delta$ of the type $\sim \mathrm{e}^{ik\eta}$. But if we consider $\vert f_{RR} \vert k^2/a^2 \gg 1$, then $\Psi$ satisfies the equation, $\Psi'' + k^2 \Psi \approx 0$, and the general solutions are of the form:
\begin{align}
\Psi = \alpha_1 \mathrm{e}^{-ik \eta} + \alpha_2 \mathrm{e}^{ik \eta},
\label{304}
\end{align}
where $\alpha_1$ and $\alpha_2$ are integration constants. Consequently we can combine Eqs. (\ref{T0}) and (\ref{Ti}) to obtain a relation between $\delta$ and $\Psi$, as $\delta '' \approx 2 \Psi ''$. Finally we can compute the general solutions of $\delta$ of the form:
\begin{align}
\delta = 2\alpha_1 \mathrm{e}^{-ik \eta} + 2\alpha_2 \mathrm{e}^{ik \eta} + \alpha_3 \eta + \alpha_4,
\label{305}
\end{align}
where $\alpha_3$ and $\alpha_4$ are integration constants. To understand the effect of magnetic field on these oscillating solutions we need to analyze these equations without approximations. Moreover, the behavior of the oscillating solutions can not be determined only by the leading order terms. We will be dealing with such issues in our future works.

\section{Evolution equations in the general relativistic case and other higher order theories}

We now derive the density perturbation evolution equation in general relativistic scenario where we can treat it as a special case of our more general result by substituting $ f(R)=0 $ in \eqref{10} and subsequent equations. Therefore, from \eqref{dense} we obtain in the GR case
\begin{align} \label{denseGR}
\left(\dfrac{\delta \rho_{m}}{\rho}\right)''+ \mathcal{H} \left(\dfrac{\delta \rho_{m}}{\rho}\right)'-\dfrac{3a^2 \Omega_{m}}{2}\dfrac{\delta \rho_{m}}{\rho}+\mathbf{O^{GR}_{B}}=0 .
\end{align}
Note that the higher curvature terms are absent in \eqref{denseGR} unlike \eqref{dense}. Consequently with appropriate approximations mentioned in the beginning of this section, equation \eqref{denseGR} can be expressed as follows:
\begin{equation}\label{delhGR}
\delta_{m} '' + \mathcal{H} \delta_{m} ' - \frac{3}{2}a^2 \Omega_{\mathrm m} \delta_{m} + \mathbf{O^{GR}_{B}} = 0,
\end{equation}
where $\mathbf{O^{GR}_{B}}  $ is given by
\begin{align}
\mathbf{O^{GR}_{B}}\equiv \dfrac{\tilde{H}}{\rho}\delta\tilde{H}''+ \dfrac{\delta\tilde{H}'}{\rho}\left[2\tilde{H}'+\left(\mathcal{H}-\dfrac{2\rho'}{\rho}\right)\tilde{H}\right]+\dfrac{\delta \tilde{H}}{\rho}\Bigg[\tilde{H}''+\left(\tilde{H}'-\dfrac{\tilde{H}\rho'}{\rho}\right)\left(\mathcal{H}-\dfrac{2\rho'}{\rho}\right)-\tilde{H}\left(\dfrac{\rho''}{\rho}+\dfrac{3a^2 \Omega_{m}}{2}\right)\Bigg].
\end{align}
Similarly in the sub-horizon approximation, equation \eqref{302} becomes
\begin{align}
\delta_{m} '' + \mathcal{H} \delta_{m} ' - 2a^2 \Omega_{\mathrm m} \delta_{m}+ \tilde{\mathbf{O}}^{GR}_{B}= 0,
\label{302GR}
\end{align}
where the magnetic contribution $\tilde{\mathbf{O}}^{GR}_{B} $ is given by
\begin{align}
\tilde{\mathbf{O}}^{GR}_{B} \equiv \dfrac{\tilde{H}}{\rho}\delta\tilde{H}''+ \dfrac{\delta\tilde{H}'}{\rho}\left[2\tilde{H}'+\left(\mathcal{H}-\dfrac{2\rho'}{\rho}\right)\tilde{H}\right]+\dfrac{\delta \tilde{H}}{\rho}\Bigg[\tilde{H}''+\left(\tilde{H}'-\dfrac{\tilde{H}\rho'}{\rho}\right)\left(\mathcal{H}-\dfrac{2\rho'}{\rho}\right)-\tilde{H}\left(\dfrac{\rho''}{\rho}+2a^2 \Omega_{m}\right)\Bigg] . \label{to:2GR}
\end{align}
We can see that in \eqref{delhGR} and \eqref{302GR} the effect of higher curvature are absent as opposed to \eqref{delh}, \eqref{302} and the equations reduce to the standard density perturbation equations with the magnetic terms denoted by $ \mathbf{O^{GR}_{B}} $ and $\tilde{\mathbf{O}}^{GR}_{B} $. We can easily check that if we switch off the magnetic field, our density perturbation equation reduces to the standard form in general relativity without any magnetic field.

Finally, it may be noted that our results can be extended to other higher order variations of gravity like $ R^{n} $ gravity and the evolution equations will change accordingly. For example, the evolution of density perturbation is then given by the following equation
\begin{equation}\label{delhRn}
\delta_{m} '' + \mathcal{H} \delta_{m} ' - \frac{3a^2 \Omega_{\mathrm m}}{2nR^{n-1}} \delta_{m} + \mathbf{O^{R^{n}}_{B}} = 0,
\end{equation}
where $\mathbf{O^{R^{n}}_{B}} $ is given by
\begin{align}
\mathbf{O^{R^{n}}_{B}}\equiv \dfrac{\tilde{H}}{\rho}\delta\tilde{H}''+ \dfrac{\delta\tilde{H}'}{\rho}\left[2\tilde{H}'+\left(\mathcal{H}-\dfrac{2\rho'}{\rho}\right)\tilde{H}\right]+\dfrac{\delta \tilde{H}}{\rho}\Bigg[\tilde{H}''+\left(\tilde{H}'-\dfrac{\tilde{H}\rho'}{\rho}\right)\left(\mathcal{H}-\dfrac{2\rho'}{\rho}\right)-\tilde{H}\left(\dfrac{\rho''}{\rho}+\dfrac{3a^2 \Omega_{m}}{2nR^{n-1}}\right)\Bigg].
\end{align}
For $ n=1$, the above equations reduce to GR. For any other value of $ n $, appropriate equations can be obtained.

\section{Discussions and Conclusions}
In this paper, we have studied the evolution of the modes of the density perturbations in the universe in presence of a cosmic magnetic field. The occurrence of primordial magnetic field (PMF) in the universe is unavoidable and it has a myriad of effects on the cosmological scenario. The stress-energy tensor of the PMF can induce additional metric perturbations, besides the primordial curvature perturbations, and also its Lorentz force affects the motion of primordial plasma. Small-scale PMFs can produce fluctuations on the baryon density field, which can affect the CMB anisotropy. Therefore, we have studied the effect of such magnetic fields on the density perturbation evolution in $ f(R) $ gravity models, i.e., in presence of dark energy or higher curvature gravitational theories, to see whether their coexistence is physically viable, in other words, whether the evolutionary consequences are realistic or not.

Here we have introduced an uniform magnetic field in order to approximate the effect of PMF. We have shown that, when the primordial magnetic field aligns with the preferred spatial direction of the LRS spacetime, it diagonalizes the shear tensor, making the shear propagation equation simpler. The effect of this PMF on the Raychaudhuri equation and other evolution equations of cosmological variables have also been analyzed. Consequently we have demonstrated the role of the PMF in structure formation. We have also found that in the subhorizon approximation, our results on scalar density perturbation conforms to that of Matsumoto \cite{Matsumoto:prd2013} in the appropriate limits, validating our results. We found that the density perturbations have an almost separate component for the magnetic field contributions in the weak field limit. Further, the more the matter dominates over the magnetic field energy contributions, the weaker is the contribution of the magnetic field perturbations in the density perturbation sector, which is a viable physical feature. This kind of approximation is possible for weak magnetic field limits, i.e., when the source of the magnetic field is weak or it is far away from the system, which is a viable approximation for the large scale study of the universe as a whole. Therefore this is not applicable to the study of local magnetized systems in $ f(R) $ gravity.

It is to be noted that the coupling between curvature and magnetic field in the action itself is very important for the generation of primordial magnetic field in $ R^{n} $ gravity, but we must clarify that in our case that we have not considered any coupling as such, i.e., we are not interested in the origin of the PMF as such but in the effect of it on density perturbation evolution in the weak field limit. In such a case, PMF is acting like any other cosmic or astrophysical magnetic field except that the strength of the field is very very weak, and that is why our approximations holds true.

This study creates the ground work for the future studies of PMF in higher order gravity theories. The robustness of the result comes from the reproducibility of the previous theoretical results. It also shows that for higher order gravities the contribution of the magnetic term increases. It is clear from our analysis that the magnetic contribution is increasing steadily from GR to $ f(R) $ and subsequently to $ R^{n} $, i.e. $ \mathbf{O^{GR}_{B}}< \mathbf{O_{B}}<\mathbf{O^{R^{n}}_{B}}$, showing increasing interplay between the magnetic field and baryons. This effect will surely get amplified further if the coupling term between curvature and magnetic field is taken into account. Consequently this will have a significant impact on structure formation, and will affect the temperature-polarization anisotropies correlation in CMBR. The effect of magnetic field energy on the expansion of the universe can not be neglected either. Therefore density perturbations in presence of PMF will affect the formation of cosmic defects. It can also influence the baryon to photon ratio and the relic neutrino temperature. Other astrophysical events like amplified effect of PMF on the scattering of dirac spinors around rotating spheroids may also occur. The presence of magnetic field will change the scattering cross section and in different gravity theories, say in $ R^{n} $ gravity, so it should be measurably different when we also consider the presence of amplified PMF in the reheating epoch.

Similar analysis can be done following the argument of \cite{Matsumoto:prd2013} for oscillating solutions in the ultra relativistic regime but such investigations will be carried out in our future work. A detailed unapproximated analysis is required for such cases. In conclusion, our work gives an approximate analysis of the density perturbation evolution in presence of a weak uniform PMF for $f(R)$ gravity and for other higher order gravity theories related to it. Our study not only extends the findings of Matsumoto on the density perturbation evolution but also demonstrates how the interplay between magnetic field and density perturbations increase with higher order curvature terms in gravity theories.

\section*{Acknowledgments}
The authors are thankful to the anonymous reviewer for the useful comments to improve the quality of the paper. SC is grateful to CSIR, Government of India for providing junior research fellowship. SC also thanks Sucheta Datta for sharing some references during the initial part of this work. SG acknowledges IUCAA, India for an associateship and CSIR, Government of India for approving the major research project No. 03(1446)/18/EMR-II. This work was conceptualized during the online program - Physics of the Early Universe - An Online Precursor (code: ICTS/peu2020/08).

\end{document}